\documentclass[preprint,12pt]{elsarticle}
\usepackage{setspace}
\usepackage{epsfig}
\usepackage{color} 
\usepackage{natbib}
\usepackage[colorlinks=true,citecolor=blue]{hyperref}
\usepackage{graphicx,amssymb,amsmath,times}
\usepackage{subfigure}
\usepackage{lscape}
\journal{New Astronomy}
\begin{document}
\begin{frontmatter}
\title{Flux and spectral variability of the blazar PKS 2155$-$304 with XMM$-$Newton:
Evidence of Particle Acceleration and Synchrotron Cooling}
\author[label1,label2]{Jai Bhagwan} 
\author[label1]{A C Gupta\corref{corr}} \ead{acgupta30@gmail.com, Tel. +91-9936683176, Fax. +91-5942-233439}
\author[label3,label4]{I. E. Papadakis}
\author[label5]{Paul J. Wiita}
\address[label1]{Aryabhatta Research Institute of Observational Sciences (ARIES), \\
Manora Peak, Nainital -- 263002, India.} 
\address [label2]{School of Studies in Physics and Astrophysics, Pt.\ Ravishankar Shukla \\
University, Amanaka G.E. Road, Raipur -- 492010, India.} 
\address [label3]{Department of Physics and Institute of Theoretical and Computational Physics, \\ 
University of Crete, GR-71003 Heraklion, Greece.}
\address [label4]{ESL, Foundation for Research and Technology, 71110 Heraklion, Greece.}
\address [label5]{Department of Physics, The College of New Jersey, PO Box 7718, Ewing, \\ 
NJ 08628-0718, USA.}

\begin{abstract}
We have analyzed {\it XMM-Newton} observations of the high energy peaked blazar, PKS 2155$-$304, made 
on 24 May 2002 in the 0.3 -- 10 keV X-ray band. These observations display a mini--flare, a nearly 
constant flux period and a strong flux increase. We performed a time-resolved spectral study of the data, 
by dividing the data into eight segments. We fitted the data with a power--law and a broken power--law model, 
and in some of the segments we found a noticeable spectral flattening of the source's spectrum below 10 keV.
 We also performed ``time-resolved" cross-correlation analyses and  detected significant hard and soft lags
(for the first time in a single observation of this source) during the first and last parts of the observation, respectively.
Our analysis of the spectra, the variations of photon-index with flux as well as the correlation and lags between the harder
and softer X-ray bands indicate that both the particle acceleration and synchrotron cooling processes make an important contribution
to the emission from this blazar. The hard lags indicate a variable acceleration process. We also estimated the magnetic field value
using the soft lags. The value of the magnetic field is consistent with the values derived from the broad-band SED modeling of this source. 
\end{abstract}

\end{frontmatter}

\section{Introduction}  
The blazar subclass of radio-loud active galactic nuclei (AGN) includes BL Lacertae objects (BL Lacs)
and flat spectrum radio quasars (FSRQs). Blazars display large amplitude flux and polarization variability 
on all possible timescales ranging from a few tens of minutes to many
years across the entire electromagnetic (EM) spectrum [1].  Blazar's center is a super massive
black hole that accretes matter and produces relativistic jets pointing almost in the direction of our line of sight 
[2]. The emission from blazars is predominantly nonthermal. The spectral energy distributions 
(SEDs) of blazars have two humps in the log$(\nu F_{\nu})$ vs log$(\nu)$ representation [3]. Based on the location 
of the low frequency hump, blazars are sub-classified into three categories: LSPs (low synchrotron peak), ISPs 
(intermediate synchrotron peak), and HSPs (high synchrotron peak) blazars [4]. The low energy 
SED hump peaks from sub-mm to soft X-ray bands and is well explained by synchrotron emission from an 
ultra-relativistic electron population residing in the magnetic fields of the approaching relativistic jet 
[5,6,7]. 
The high energy SED hump that peaks in the MeV--TeV gamma-ray bands is not as well understood but is usually 
believed to be attributed to  inverse Compton (IC) scattering of  photons off those relativistic electrons. 

Variability in blazars on timescales of a few minutes to less than a day often is known as intra-day variability
(IDV) [8]; variability timescales from  days to several weeks is sometimes called short timescale
variability (STV), while variability on month to  year timescales is known as long term variability (LTV) [9]. 

X-ray variability in HBLs is characterized by  correlated changes of the spectral index with the 
X-ray flux and  lags between soft and hard X-rays. A photon index-flux correlation was first observed by 
[10] in Mrk 421 using {\it EXOSAT}.  Similar results were also obtained for other HBLs [11,12]. 
A soft lag in the HBL H 0323$+$022 in {\it Ginga} X-ray observations was observed
for the first time by [13]. 

PKS 2155$-$304 ($\alpha_{2000.0} =$ 21h 58m 52.0s, $\delta_{2000.0} = $-30$^{\circ}$ 13$^{\prime}$ 
32$^{\prime \prime}$) at $z = 0.116$ 
is a HSP blazar that is highly variable across the entire EM spectrum [14].
It is the brightest blazar in the UV to $\gamma-$ray bands in the southern hemisphere. 
Blazar variability is best studied throughout different phases which can be considered to be: outburst, 
pre/post outburst, and  low states. PKS 2155-304 is one of the most commonly observed blazars for simultaneous 
multi-wavelength observations and has received maximum attention for simultaneous multi-wavelength campaigns  
[15$-$33]. This blazar has also been observed in assorted single EM bands over more diverse timescales. For instance, [34] have studied 
the long term optical variability of this source while [14] have studied the X-ray IDV in the 
blazar. There is some evidence for PKS 2155$-$304 having shown quasi-periodic oscillations (QPOs) on IDV time 
scales from {\it IUE} and {\it XMM-Newton} observations [14,16,35]. 
[36,37]  observed PKS 2155$-$304 in the high flux state and detected  
pronounced spectral variations. Using {\it XMM-Newton} data [38] have reported that the synchrotron emission of 
PKS 2155$-$304 then peaked in UV$-$EUV bands rather then the soft X-ray band. [39] fitted 
the SED of PKS 2155$-$304 with a log-parabolic model and found that the curvature of the SED is anti-correlated 
with the peak energy $E_{\rm p}$.   

In a recent paper [40], we used all the archival XMM/Newton observations of PKS 2155 - 304 in 
order to study its broad band (optical/UV/X--rays) flux and spectral variability. We found that the long term 
optical/UV and X-ray flux variations in this source are mainly driven by model normalization variations. We also 
found that the X-ray band flux is affected by spectral variations. Overall, the energy at which the emitted power 
is maximum correlates positively with the total flux. As the spectrum shifts to higher frequencies, the spectral 
`curvature' increases as well, in contrast to what is expected if a single log-parabolic model were an acceptable 
representation of the broad band SEDs. These results suggested that the optical/UV and X-ray emissions in this 
source may arise from different lepton populations. 

We have now started the study of the individual XMM/Newton observations of the same source, with the aim to 
study its short term X--ray variability properties, since these data provide an excellent opportunity 
to analyze and model the blazar emission in different flux states with identical instrumentation. In this paper, 
we present the results from the analysis of the XMM/Newton observation of PKS 2155-304 which was taken on 24 May 2002. 
The same data have been analyzed by [27,28]. They have studied the short term variability and 
cross correlation analysis between the different X-ray energy bands. The observation includes three nearly equal 
length pointings at the blazar with a total exposure time of 93 ks. But each exposure has been taken with different 
filters. We noticed that the combined light curve of these three pointings shows nearly stable flux states, 
declining flux states, rapid flares, and weak oscillations. We have found evidence for flux related spectral variations
(which is typical of this source), but we also find evidence (for first time) for the presence of both ``hard" and ``soft" time lags,
which are variable with time. Since the flux variability behavior in this observation is rather typical of the source, we believe that the
results we present in this paper may be representative of the X--ray 
variability properties of the source. This will be confirmed when we will have finished the analysis of the 
remaining observations as well. The final results will hopefully offer us a more complete view on the physical 
processes which dominate the X-ray flux and spectral evolution in the source.  

The paper is structured as follows. In section 2, we present the data and our reduction procedure.  In section 3 we report the results,
and in section 4 we discuss them in the context of acceleration and/or cooling processes that may operate in the system, and we present our summary.

\section{Data and Reduction}
We analyzed the archival XMM-Newton EPIC/pn data of the blazar PKS 2155$-$304 in the 0.3 -- 10 keV 
X-ray band. These observations were made on 24 May 2002 (orbit 450, Obs ID 0124930501; PI: Fred Jansen). 
This observation ID has three continuous EPIC/pn exposures in small window (SW) mode with different filters: 450-1 
was taken in the medium filter, 450-2 in the thin filter and 450-3 in the thick filter. The duration of the respective cleaned 
files are 31.7, 31.6 and 29.7 ks, respectively. The Original Data File (ODF) was reprocessed  using 
Science Analysis System (SAS) version 11.0.0 with the most recent available calibration files. 

By generating a hard band background light curve in the energy range 10--12 keV, we have checked 
for the high soft proton background periods which are caused by solar activity. We  removed those points 
for which the hard band count rate was greater than 0.4 count/sec, and then we generated the good time 
interval (GTI) data. We have used the single and double events with quality and pattern flags constrained 
to $(FLAG=0) \& (PATTERN \leq 4)$ for our analysis. We have carefully examined the pile-up effect in the 
data by SAS task {\it epatplot} and found that the data is indeed affected by pile-up.  As a result, we 
extracted the source count from an annuls region which was centered on the source with inner radius of 
$10^{\prime \prime}$ and outer radius of $40^{\prime \prime}$. The background counts were accumulated 
from a circular region of radius $45^{\prime \prime}$ on the CCD chip near where the source was 
located and least affected from the source counts.

 We extracted background subtracted light curves by using the $epiclccorr$ 
command. The spectral analysis was done by using the {\it xmmselect} and {\it specgroup} tasks in SAS.  We have used the 
task {\it specgroup} to rebin all spectra, in order to have at least 50 counts for each 
background subtracted spectral channel and the value of over sampling parameter is taken as 5. 

\section{Results}
\subsection{Light Curves}
 We present the light curves extracted from entire observation of PKS 2155$-$304 in the five energy  bands of 0.3 - 0.5 keV,
 0.5 - 2.0 keV, 2.0 - 4.0 keV, 4.0 - 10.0 keV and 0.3 - 10.0 keV (with an 100 s binning)  in Fig.\ 1. As mentioned above,
 this observation consists 
of three exposures with 
a duration 31.7, 31.6 and 29.7 ks, respectively. The gaps between the individual exposures are of the order of $\sim 1.5$ ks. 
The three sub-parts are easily spotted by the abrupt change in the source flux. For example, the flux drop at around $\sim 65$ ks since 
the start of the observation is not real, but is due to the change of the EPIC/pn filters, since the EPIC/pn thin filter is more
transparent to the soft energy photons than the thick filter. A similar effect, but of a much smaller amplitude, is observed $\sim 30$ ks
since the start of the observation, when the EPIC/pn filter was switched from medium to thin. 

Despite the EPIC/pn filter changes, a visual inspection indicates significant and clearly defined flux variability in the light curve.
In Fig.\ 1, the individual exposures are marked as segments 1, 2, and 3.  We divided segment 1 into 
three sub-segments: 1(a) where the source flux is almost constant; 1(b) where the flux decreases; and 1(c) where the source flux rises.
Segment 2 is relatively flat, but this is when the source has shown a hint of a QPO [14]. Segment 3 is divided into 
four sub-segments: 3(a), with decreasing flux, 3(d) corresponds to the $\sim$ constant flux level towards the end of the observation,
while 3(b) and 3(c) are defined during the first and second part of the strong rising flux state in between.
The percentage variability and signal to noise ratio (S/N) of segment 1, 2 and 3 in 0.3 -- 10.0 keV band are 6.6$\pm$0.16, 1.5$\pm$0.22,
21$\pm$0.15 and 37.7, 39.3, 38.0, respectively.

\subsection{Spectral Variability}
To investigate if there are spectral variations associated with the well defined flux variations, we generated the spectra of the 
eight sub-segments we discussed above. We then fitted a simple power law (PL) model to the X-ray spectra using {\tt XSPEC} (v. 12.8.0).
The average value of Galactic 
neutral hydrogen absorption was found from the HEASARC N$_H$ calculator to be
$ N_{H} = 1.71 \times 10^{20}$ cm$^{-2}$ [41]. We kept this value fixed in all spectral fits. 
The best model fit results (reduced $\chi^{2}$ values and model parameters with their corresponding 90\% confidence intervals)  
are summarized in  Table 1. In the last three columns of the same Table we list the best-fit flux estimates (together with
their 90\% confidence limits) in the 0.3-2, 2-10 and 0.3-10 keV bands. The best fit PL models, and ratios of data to model are
plotted in Fig.\, 2 for all segments.
Based on the null probability for the PL model fits (listed in the 7th column of the upper part of Table 1) the model provides
statistically acceptable fits to the spectra of all segments. The exception are the spectral of segments 3(b) and 3(d), in particular,
where the probability is low. The data-to-model ratios for segments 3(d)(bottom right panels in Fig. 2) show a convex shape
above $\sim 2$ or $3$ keV. This shape is indicative of a more complex spectral shape.

For that reason, we also fitted the spectra with a broken power-law model (BKPL). The best BKPL model fit results are also listed in Table 1.
The model resulted in entirely unconstrained break-energies, and almost identical $\Gamma_1$ and $\Gamma_2$ best fit values, in the case of
segments 1(a), 1(b), 1(c). These results imply that a single PL model is adequate for the characterization of the energy spectra of these
segments. For this reason, we do not list the BKPL best-fit results for these segments in Table~1. Contrary to this, the BKPL
best-fit $\chi^2$ values are smaller than the respective PL best-fit values, for the remaining segments. The numbers in parenthesis in
the 7th column of Table 1 indicate the probability of the $F-$statistic given the $\chi^2$ values of the PL and BKPL best-fits. Based on
the $F-$ratio probabilities, the improvement in the fit when we consider the BKPL model is significant at more than the 99\% level in
the case of segment 2 and all sub-segments of the last part of the 
observation. 

Our results indicate that during segments 2 and 3(a), when the flux remains rather constant, the spectra steepen above the break energy
by an average $\Delta\Gamma=0.11\pm 0.08$. A stronger spectral steepening, by a factor of $\Delta\Gamma=0.55\pm0.27$ is also
observed during segment 3(d), when the flux remains again $\sim$ constant, albeit at a higher than previous level. In all these three cases,
the break-energy is rather high, with an average value of $3.4\pm 0.7$ keV. On the contrary, the strong flux rise observed during segments
3b and 3c, is associated with a small, but significant, spectral hardening at high energies, of $\Delta\Gamma=-0.07\pm0.02$, between the
spectral slope below and above the break-energy. At the same time, the average break energy for these segments is $1.26\pm0.32$, which is
significantly smaller than the break-energy in the spectra of segment 3d.

``Log-parabolic" models have also been used to describe 
continuously downward-curving spectra of HSPs [42]. We fitted the spectra of all segments with such a model, but the quality of the fits
is not significantly better than that of the PL fits. In fact, in some cases, the $\chi^2$ values are even larger than those from the PL
fits (for a smaller number of dof). The best-fit curvature parameter turns out to be negative only for segments 1(b) and 1(c), but even in
this case, the values are  consistent with zero within the errors.

Using the PL best-fit results, we investigated whether the spectral variations are related with the flux variations of the source.
Figure 3 shows a plot of the spectral PL index{\footnote{Note that the plot in  Fig.\,3 does not change if we use the best-fit $\Gamma_1$ values
from the BKPL fits, as they are almost identical to the best-fit $\Gamma$ values of the PL fits.}as a function of the source 2--10 keV flux.
In general, we find that the spectral slope is anti-correlated 
with flux; the spectra ``harden" with increasing flux. Using the data plotted in Fig.\ 3, we find a correlation coefficient of $r = -0.95$,
which is significant at better than the 99\% confidence level  ($P_{null} = 3.0\times 10^{-4}$). Looking the data in more detail,
the source flux and spectral index are similar for the first three segments (1a, 1b and 1c), while the spectral index flattens as the
2--10 keV flux increases during segments (2) and (3a). Taken as a whole, the spectral-flux evolution from (1a) to (3) may exhibit a ``clockwise
trajectory" in the ``spectral slope -- flux" plane, although the flux and spectral variations are not of very large amplitude.
The spectrum shows a strong ``hardening" during the strong flux rise towards the end of the observation. 

 \subsection{Correlation Analysis}

The spectral variations reported in the previous section suggest possible energy delays in the observed flux variations. In order to
investigate further the spectral variability of the source, we  estimated the Cross-Correlation Function (CCF) between light curves at
different energy bands, using light curves with a 100 s bin size. We chose light curves in the energy bands 0.3--0.5, 0.5--2, 2--4 and
4--10 keV. We calculated the sample CCF as in Section 3 of [43], and we estimated the CCF between the 4--10, 2--4, and 0.5--2 keV band
light curves versus the 0.3--0.5 keV band light curve. Significant correlation at positive lags means that soft (i.e. the 0.3--0.5 keV)
band variations are leading the variations in the ``harder" band. 

We followed a similar, but simplified, version of the ``sliding window " technique of [43]. We  calculated the CCFs for 10 ks intervals,
which started at different times of the observed light curves. More specifically, we used 10 ks data streams which started at the points
$t_{\rm start}=0, 500, 1000, \ldots$ s since the start of the observation. As a result, we computed the CCF for $\sim 170$ 10 ks long
segments for the complete light curve. For each CCF, we identified the lag,  $\tau_{\rm max}$, at which the CCF was maximum (i.e. CCF$_{\rm max}$),
and calculated the mean time lag, $\tau_{\rm mean}$,  as the mean of the time lags of the points with CCF values larger or equal
to 0.8$\times$CCF$_{\rm max}$, in the case when: a) the CCF$_{\rm max}$ value was larger or equal to 0.8, and b) there were at least
five points with a CCF value larger than 0.8 of CCF$_{\rm max}$. In other words, we have only considered  CCF peaks which are strong,
broad and well defined, as opposed to random, narrow peaks in the CCF. 
Finally, we did not consider light curve segments in the time intervals between 21.7-43 ks and 54.6-76 ks after the start of the observation,
as in this case, the light curves would include data during the EPIC/pn filter switches (which affect in a different way the observed count
rate in the soft and harder energy bands). 

The above procedure allows us to monitor the temporal variation of the time lag between various bands. Our results are plotted in Fig.\, 4.
The top panel in this figure shows the 4--10 keV band light curve (which is the most representative of the intrinsic source variations,
as this band is less affected by the filter changes). The bottom panel shows the average time lag between the 0.5--2, 2--4 and 4--10 keV band
light curves and the 0.3--0.5 keV band light curve (red filled circles, blue open circles, and black filled circles, respectively). The vertical
dashed lines in the top panel indicate the light curve segments that were used to estimate the mean time lags plotted in the bottom panel.
In addition to the intervals that we excluded for the estimation of the CCF, the resulting CCF for the intervals $\sim 17-22$ ks and $43-55$ ks
did not show any significant peaks in them. This is most probably due to the fact that all light curves were relative flat during these periods,
with very low amplitude 
variations, roughly similar to the Poisson noise variations (specially in the high energy bands). Consequently, the sample CCFs are not capable
to detect any intrinsic delays in these periods.

Our results indicate that the correlation between the ``soft" and ``hard" bands is rather complex. In the first part of the observation
(the part which corresponds to segments (1a), (1b) and the first part of segment (1c) the lags are {\it positive}, i.e. the variations in
the 0.3--0.5 keV band are {\it leading} those in the harder bands (i.e. we observe a {\it hard lag}). In this case, whatever the characteristic
time scales which dominate the observed variations, they must be faster at low energies. Furthermore, the time lags between the highest energy
bands (4--10 and 2--10 keV) and 0.3--0.5 keV appear to decrease with time, from $\sim 1.5$ to $\sim 0$ ks. At the same time, the 0.5--2 vs
0.3--0.5 keV lags appear to be constant at $\sim 100$ s. 

The reverse situation is observed in the last part of the observation, $\sim 80$ ks after the start of the observation, during the continuous,
large amplitude flux rise. During the flux rise, the time delays between 0.5--2 and 0.3--0.5 keV, as well as between the 2--4 and 0.3--0.5 keV
bands are consistent with zero, while the 4--10 keV band variations may be {\it leading} the soft band variations. When the source flux flattens
after $\sim 85$ ks, the harder band variations are clearly {\it leading} the soft band variations (i.e. we observe a {\it soft lag}). The delay
is energy dependent: the 0.3--0.5 keV band variations are delayed with respect to the 4--10 and 2--4 keV band variations by $\sim 700$ s, while
the time lag between the 0.5--2 and the 0.3--0.5 keV bands is $\sim 300$ s . }

curve and the 0.5--2, 2--4 and 4-10 keV band light curves, in the bottom panel (red filled circles, blue open circles and black field circles,
respectively).


curve 
respectively).

\section{Discussion and Conclusions}

We have studied the flux and spectral variability of PKS 2155--304 using a $\sim 100$ ks XMM/Newton observation. The flux variability detected
in this observation is rather typical of this source; we observe a short flare, decaying and stable flux states, as well as a rapid,
large amplitude flux increase towards the end of the observation. We performed a time-resolved spectroscopic analysis, by dividing the
data in various segments with different mean flux, and we fitted a simple power-law model to them.

\subsection{The spectral variability results}In general,  X-ray spectra of HSPs are rather steep 
$(\Gamma ~ \textgreater ~ 2)$ and continuously steepen with increasing energies [44,45]). As a result, X-ray spectra of HSPs follow a convex,
or continuously downward-curved, shape. This 
is a signature of the high energy tail of synchrotron emission that is the outcome of energy dependent 
particle acceleration and cooling [42]. In the last decade, X-ray observations 
of PKS 2155--304 by various X-ray telescopes have shown that its spectra normally has a convex shape below 
10 keV [39,46-49]. 
However, two observations by {\it XMM-Newton} of this blazar in 2006  were analyzed by [38] and he
reported that the shape of those X-ray spectra were concave. 

Our observations indicate that a simple power-law model can describe adequately the 
source's spectra during the first part of the observation, when we observe a mini-flare. In the rest of the observation, our analysis
indicate that the energy spectra are more complex. We observe a spectral steepening above $\sim 3.4$ keV, by a factor of
 $\Delta\Gamma\sim 0.1-0.5$ during segments 2, 3(a) and 3(d), when the flux remains rather constant.

 In addition, during the flux rise detected towards the end of the observation (segments 3b and 3c) the X-ray spectra show the opposite behavior;
 we observe a spectral hardening, of low amplitude ($\Delta\Gamma\sim 0.1$, above $\sim 1-1.5$ keV. This could be an indication of the low-energy
 side of the Inverse Compton emission contributing to energies even below 10 keV. A detailed, similar analysis of all the available XMM-Newton
 observations of the source, could provide useful information regarding the spectral hardening/softening (amplitude of $\Delta\Gamma$ and
 location of $E_{\rm break}$) and its relation to the flux variations of the source. 

When we plotted the spectral index as a function of flux, we found the typical ``flatter when brighter" spectral variability behavior,
in agreement with previous studies [28,32,48]. This anti-correlation between spectral slope and source flux  is a signature of spectral
flattening with increasing flux, and indicates that the hard X-ray band flux increases more than the soft X-ray band flux as the flux increases.
There is an indication of spectral variations along a ``clockwise" trajectory on the spectral slope -- flux plane, but the evidence is not very
strong. 

\subsection{The cross-correlation results}
We also performed a detailed cross-correlation analysis of the variations detected, by dividing the light curve into 4 energy bands.
We detected both hard lags, as well as soft lags, in the first and last part of the observation, respectively. To the best of our knowledge,
this is the first time that both hard and soft lags are detected in a light curve of this source. 

In the simplest case scenario, the X-ray emission variation of HSPs is controlled by the energy dependent particle acceleration and cooling 
mechanisms. The cooling process is well understood but particle acceleration is not well understood yet and 
could operate in different ways [50,51]. Following [51], the acceleration timescale, $t_{\rm acc}$, of the relativistic electrons, and
their cooling time scale, $t_{\rm cool}$, in the observed frame, can be expressed as a
function of the observed photon energy E (in keV) as follows: 

\begin{equation}
t_{\rm acc}(E)= 9.65\times10^{-2}(1+z)^{3/2}\xi B^{-3/2}\delta^{-3/2}E^{1/2} {\rm s},
\end{equation}
\noindent
and,
\begin{equation}
t_{\rm cool}(E)= 3.04\times10^{3}(1+z)^{1/2}B^{-3/2}\delta^{-1/2}E^{-1/2} {\rm s},
\end{equation}
\noindent
where $z$ is the source red shift, $B$ is the magnetic field (in Gauss), $\delta$ is the Doppler factor of 
the emitting region, and $\xi$ is a parameter describing how fast the electron can be accelerated. 

Equations (1) and (2) show that both $t_{\rm cool}$ and $t_{\rm acc}$ depend on the observed photon 
energy but do so in an inverse fashion. The lower energy electrons are radiating low energy photons which 
cool slower but are accelerated faster than the high energy electrons that are radiating the high energy 
photons. The relationship between  $t_{\rm cool}$ and $t_{\rm acc}$ can in principle give an important clue about the 
process that dominates the X-ray emission. 

If $t_{\rm cool}$ is significantly larger than $t_{\rm acc}$, the cooling process dominates [52]. In this case, any change in
emission will propagate from higher to lower energies so that 
higher energy photons will lead lower energy photons (soft lag) and a clockwise loop of spectral 
index against the flux will be observed. The expected soft lag should be approximately equal to
\begin{equation}
\tau_{\rm soft}= t_{\rm cool}(E_{\rm l})-t_{\rm cool}(E_{\rm h}) , 
\end{equation}
\noindent
where $\tau_{\rm soft}$ is the observed soft lag, and the values of $E_{\rm l}$ and  $E_{\rm h}$ are the 
energy of the lower and higher energy bands in units of keV, respectively. 

In contrast, if  $t_{\rm acc}$ is comparable to the $t_{\rm cool}$ in the observed energy range, the system is dominated by
acceleration processes and any variation in 
emission propagates from lower to higher energies because it takes a longer 
time to accelerate a particle to higher energies. In this case there would be a hard lag and an 
anti-clockwise loop would be observed in spectral index versus flux plot. The  time lag in an acceleration dominated system is 
expressed as,
\begin{equation}
\tau_{\rm hard}= t_{\rm acc}(E_{\rm h})-t_{\rm acc}(E_{\rm l})  ,
\end{equation}
\noindent
where  $\tau_{\rm hard}$ is the observed  hard lag  between the low and high energy bands, respectively.

\subsection{Implications of the hard lags}Based on this scenario, the hard lag we observe in the first part of the observation may
indicate that the system is dominated by acceleration processes. We observe that $\tau_{hard}(4-10 {\rm vs} 0.3-0.5) > \tau_{hard}(2-4 {\rm vs}
0.3-0.5) > \tau_{hard}(0.5-2 {\rm vs} 0.3--0.5)$, which is easy to understand since the energy separation of these bands increases accordingly.
Using equations (1) and (4), we can show that, in this case, we should expect that

\begin{eqnarray}
B\delta\xi^{-2/3}= 0.21(1+z)E_{h}^{1/3}\left[\frac{1-(E_{l}/E_{h})^{1/2}}{\tau_{\rm hard}}\right]^{2/3}.
\end{eqnarray}
\noindent
We found that $\tau_{hard}(4-10 {\rm vs} 0.3-0.5)$ and $\tau_{hard}(2-4 {\rm vs} 0.3-0.5)$ do not remain constant, but rather decrease with
time. According to eq.\,(5), this would imply that either $B$ and/or $\delta$ {\it increase} with time, or that the parameter $\xi$ {\it decreases}
with time. We also observed that $\tau_{hard}(0.5-2 {\rm vs} 0.3-0.5)$ to remain constant in the same period. This result indicates that neither
$B$ nor $\delta$ should control the variations with time of the $\tau_{hard}$ in the higher energy bands, as in this case we would expect these
lags to decrease with time.

One possibility is that the acceleration process dominates the variations above 2 keV, but at lower energies a different variability
process dominates the observed variations. This is though highly unlikely, given the very good correlation between the variations below
and above 2 keV. Perhaps then,the variability in the first part of the observation is dominated by acceleration process, and that the
rate of acceleration, $\xi$, does not remain constant, but rather decreases with time. In addition, if a variable $\xi$ determines the
variability evolution of the system then, apart from being time dependent, it may also by energy dependent, in which case the constant
low-energy time lags could also be explained.

The probability of the system to be ``acceleration dominated" contradicts the spectral slope - flux evolution in the first part of
the observation. Our 
spectral analysis suggests a clockwise trajectory in the spectral index versus flux plot during the first part of the observation.
Arguably the amplitude of this ``trajectory" is not very strong, but it is fair to notice that we find no evidence of an ``anti-clockwise"
evolution, which is what we would expect if the acceleration process dominates the variability evolution of the source. We cannot offer an
interpretation at the moment, as to why could be the reason for these seemingly contradictory results. 

\subsection{Implications of the soft lags} On the other hand, the soft lags we observe in the last part of the observation suggest that
the system is dominated by cooling processes. We observe again that
$\tau_{hard}(4-10 {\rm vs} 0.3-0.5) > \tau_{hard}(2-4 {\rm vs} 0.3-0.5) > \tau_{hard}(0.5-2 {\rm vs} 0.3-0.5)$ (in absolute value),
which can be again understood due to the accordingly increasing energy separation of these bands. Using equations (2) and (3), we
expect that, in this case,  

\begin{eqnarray}
 B\delta^{1/3}=209.91\left(\frac{1+z}{E_{l}}\right)^{1/3}\left[\frac{1-(E_{l}/E_{h})^{1/2}}{\tau_{soft}}\right]^{2/3}.
\end{eqnarray}
\noindent
The above equation indicates that the observed time lags should be inversely 
proportional either to the magnetic field $B$ of the emitting blob ($B \propto \tau^{-2/3}$) and/or  to $\delta$ ($\delta \propto\tau^{-2/3}$).
In the last part of the observation, we observe that all soft lags remain roughly constant. Their (absolute) mean values at times longer than
85 ks (since the start of the observation)
are: $<\tau_{soft}(4-10 {\rm vs} 0.3-0.5)>=726\pm 54$ s, $<\tau_{soft}(2-4{\rm vs}0.3-0.5)>=625\pm 25$ s, and $<\tau_{soft}(0.5-2{\rm vs}0.3-0.5)>=317\pm 36$ s.
Assuming that $\delta t=10$ and that it remains constant during the observation, then the above values together with eq.\,(6),
result in the following estimates for the magnetic field strength: $1.8\pm 0.2$ G, $1.5\pm 0.05$ G, and $1.5\pm 0.1$ G.
The weighted mean value is $1.5\pm 0.05$ G, which is similar to the magnetic field values that result from the broad band SED
fitting of blazars [53]. We note that, according to our results, $\tau_{soft}$ is much smaller by a factor of $\sim 5-6$
(for all energy bands) during the flux 
rise (at times $\sim 75-85$ ks). This would imply an {\it increased} magnetic field value, by a factor of $\sim 3$ during the flux rise phase. \\
(28) have analyzed the same observations and done cross correlation analysis in which they divided the each exposure light curve in
three energy (e.g., 0.2-0.8 keV (soft), 0.8.2.4 keV (medium) and 2.4-10.0 keV (hard)) bands. In their analysis they have calculated the
time lags between soft/medium and soft/hard energy bands. They have found the soft band variations lag behind the medium and hard
band by $\sim 450\pm300$, $\sim 870\pm540$  and $\sim 420\pm180$, $\sim 1280\pm240$ seconds, respectively in segment 1 and 3 of
the light curves. Our results are comparable to these results in segment 3 and opposite in segment 1.

\subsection{Summary}
Our analysis suggest the presence of  soft lags in some epochs and hard lags in other 
epochs, during the May 2002 XMM-Newton observation of PKS 2155-304. This result implies that the difference
between $t_{\rm acc}$ and $t_{\rm cool}$ of the emitting electrons is changing 
from epoch to epoch, in agreement with past studies [28,48,49]. 

  Hard time lags are detected during the first part of the observation. These lags suggest that the flux evolution during this part
 of the observation is dominated by the acceleration process. These delays may be  modulated by variations of the acceleration
 parameter $\xi$ in time, and perhaps in energy as well).  The shock formation process due to collision is the 
most acceptable reason for particle acceleration [54,55]. If real, the variation in the 
value of $\xi$ indicates that the particle acceleration mechanisms 
may be time variable. Using the soft time lags observed in the last part of the observation we have estimated the value of
the magnetic field, which turns out to be similar to the values that have been reported in the literature, based on the
modeling of the full band SEDs of blazars. We also found evidence for the magnetic field variations, during the flux
rise phases in the light curve.
    
The X-ray observations are the most powerful diagnostic tool for blazars which have an ability to provide 
information about the physical processes taking place in the vicinity of the central engines of these sources. 
We believe that our results demonstrate that, in order understand the relationship between particle acceleration and cooling, 
combined spectral and timing studies within the X--ray band can be useful. Future, similar studies of the archival XMM/Newton
observations of this source may allow us to understand better the physical processes that control the flux and spectral variations in this source. 

\section*{Acknowledgments}
We thank the referee for detailed and thoughtful comments which has helped us to improve the manuscript.
This work is based on the observations obtained with XMM-Newton, an ESA science mission with instruments and 
contributions directly funded by the ESA member states and NASA. 




\newpage
\begin{table*}
{\bf Table 1. Results of X-ray spectral fits to PKS 2155$-$304} \\ \\ 
\scriptsize 
\hspace*{-0.5in}
\begin{tabular}{llllcccrrr}  \hline  
Segment & $\Gamma_1$ &  N$^{a}$ & E$_{\rm break}$ & $\Gamma_{2}$ & ${\chi^2}$/dof & Prob. & Flux$^{b}$ & Flux$^{c}$ & Flux$^{d}$   \\ 
\hline 
\multicolumn{10}{|c|}{Power-Law (PL) Model} \\
\hline 
\\
1(a) & 2.66$_{-0.01}^{+0.01}$  & 3.04$_{-0.02}^{+0.02}$ &... &... & 151.8/142 & 0.27 & 1.010$_{-0.004}^{+ 0.003}$ & 0.305$_{-0.003}^{+0.003}$ & 1.315$_{-0.005}^{+0.004}$ \\ 
1(b) & 2.70$_{-0.01}^{+0.01}$  & 2.65$_{-0.01}^{+0.01}$ &... &... & 157.1/157 & 0.48 & 0.898$_{-0.002}^{+0.002}$  & 0.250$_{-0.002}^{+0.002}$ & 1.148$_{-0.003}^{+0.003}$  \\ 
1(c) & 2.68$_{-0.01}^{+0.01}$  & 2.63$_{-0.01}^{+0.02}$ &... &... & 185.0/177 & 0.34 & 0.882$_{-0.002}^{+0.002}$  & 0.255$_{-0.002}^{+0.002}$ & 1.137$_{-0.003}^{+0.003}$  \\ 
2    & 2.58$_{-0.01}^{+0.01}$  & 2.86$_{-0.01}^{+0.01}$ &... &... & 222.5/207 & 0.22 & 0.921$_{-0.001}^{+0.001}$  & 0.320$_{-0.001}^{+0.001}$ & 1.241$_{-0.002}^{+0.002}$ \\ 
3(a) & 2.56$_{-0.01}^{+0.01}$  & 3.04$_{-0.02}^{+0.02}$ &... &... & 172.4/156 & 0.17 & 0.973$_{-0.003}^{+0.004}$  & 0.347$_{-0.003}^{+0.003}$ & 1.321$_{-0.005}^{+0.004}$  \\ 
3(b) & 2.46$_{-0.01}^{+0.01}$  & 3.16$_{-0.02}^{+0.02}$ &... &... & 205.8/174 & 0.05 & 0.976$_{-0.004}^{+0.004}$  & 0.420$_{-0.003}^{+0.003}$ & 1.395$_{-0.004}^{+0.005}$  \\ 
3(c) & 2.32$_{-0.01}^{+0.01}$  & 4.12$_{-0.02}^{+0.02}$ &... &... & 182.9/178 & 0.38 & 1.217$_{-0.004}^{+0.004}$  & 0.676$_{-0.005}^{+0.006}$ & 1.885$_{-0.007}^{+0.007}$  \\
3(d) & 2.32$_{-0.01}^{+0.01}$  & 4.89$_{-0.02}^{+0.02}$ &... &... & 247.3/179 & 5.5e-4 & 1.444$_{-0.005}^{+0.004}$  & 0.793$_{-0.006}^{+0.006}$ & 2.236$_{-0.005}^{+0.007}$  \\\\
\hline 
\multicolumn{10}{|c|}{Broken-Power-Law (BKPL) Model} \\
\hline 
\\
2    & 2.58$_{-0.01}^{+0.01}$  & 2.86$_{-0.01}^{+0.01}$ &3.27$_{-0.78}^{+2.16}$ & 2.65$_{-0.04}^{+0.17}$ & 210.7/205 & 0.38 & 0.921$_{-0.001}^{+0.002}$  & 0.315$_{-0.002}^{+0.002}$ & 1.236$_{-0.002}^{+0.003}$  \\ 
& & & & & & (4.0e-3) & & & \\
3(a) & 2.55$_{-0.02}^{+0.01}$  & 3.05$_{-0.02}^{+0.02}$ &2.95$_{-1.50}^{+0.83}$ & 2.69$_{-0.09}^{+0.10}$ & 160.3/154 & 0.35 & 0.972$_{-0.003}^{+0.003}$  & 0.338$_{-0.004}^{+0.005}$ & 1.310$_{-0.004}^{+0.007}$  \\ 
& & & & & & 4.0e-3 & & & \\
3(b) & 2.49$_{-0.02}^{+0.02}$  & 3.13$_{-0.03}^{+0.02}$ &1.26$_{-0.25}^{+0.60}$ & 2.42$_{-0.03}^{+0.02}$ & 190.4/172 & 0.16 & 0.977$_{-0.003}^{+0.004}$  & 0.431$_{-0.004}^{+0.003}$ & 1.408$_{-0.005}^{+0.007}$  \\ 
& & & & & & (1.2e-3) & & & \\
3(c) & 2.35$_{-0.02}^{+0.03}$  & 4.08$_{-0.07}^{+0.03}$ &1.26$_{-0.44}^{+0.50}$ & 2.28$_{-0.02}^{+0.02}$ & 170.2/176 & 0.61 & 1.220$_{-0.005}^{+0.005}$  & 0.683$_{-0.006}^{+0.009}$ & 1.902$_{-0.007}^{+0.011}$  \\ 
& & & & & & (1.8e-3) & & & \\
3(d) & 2.30$_{-0.01}^{+0.01}$  & 4.90$_{-0.02}^{+0.02}$ &4.93$_{-1.15}^{+0.47}$ & 2.85$_{-0.31}^{+0.22}$ & 196.3/177 & 0.15 & 1.442$_{-0.004}^{+0.005}$  & 0.761$_{-0.007}^{+0.008}$ & 2.203$_{-0.008}^{+0.010}$  \\ 
& & & & & & (1.3e-9) & & & \\
\hline  
\end{tabular} \\
$^a$ Normalization constant in units of 10$^{-2}$  photon cm$^{-2}$s$^{-1}$keV$^{-1}$ \\
$^b$ Unabsorbed 0.3--2.0 keV model flux in units of 10$^{-10}$ erg cm$^{-2}$ s$^{-1}$ \\
$^c$ Unabsorbed 2.0--10.0 keV model flux in units of 10$^{-10}$ erg cm$^{-2}$ s$^{-1}$ \\
$^d$ Unabsorbed 0.3--10.0 keV model flux in units of 10$^{-10}$ erg cm$^{-2}$ s$^{-1}$ \\ 
\end{table*} 

\newpage
\begin{figure}
\begin{center}
\includegraphics[width=1.2\textwidth,angle=0]{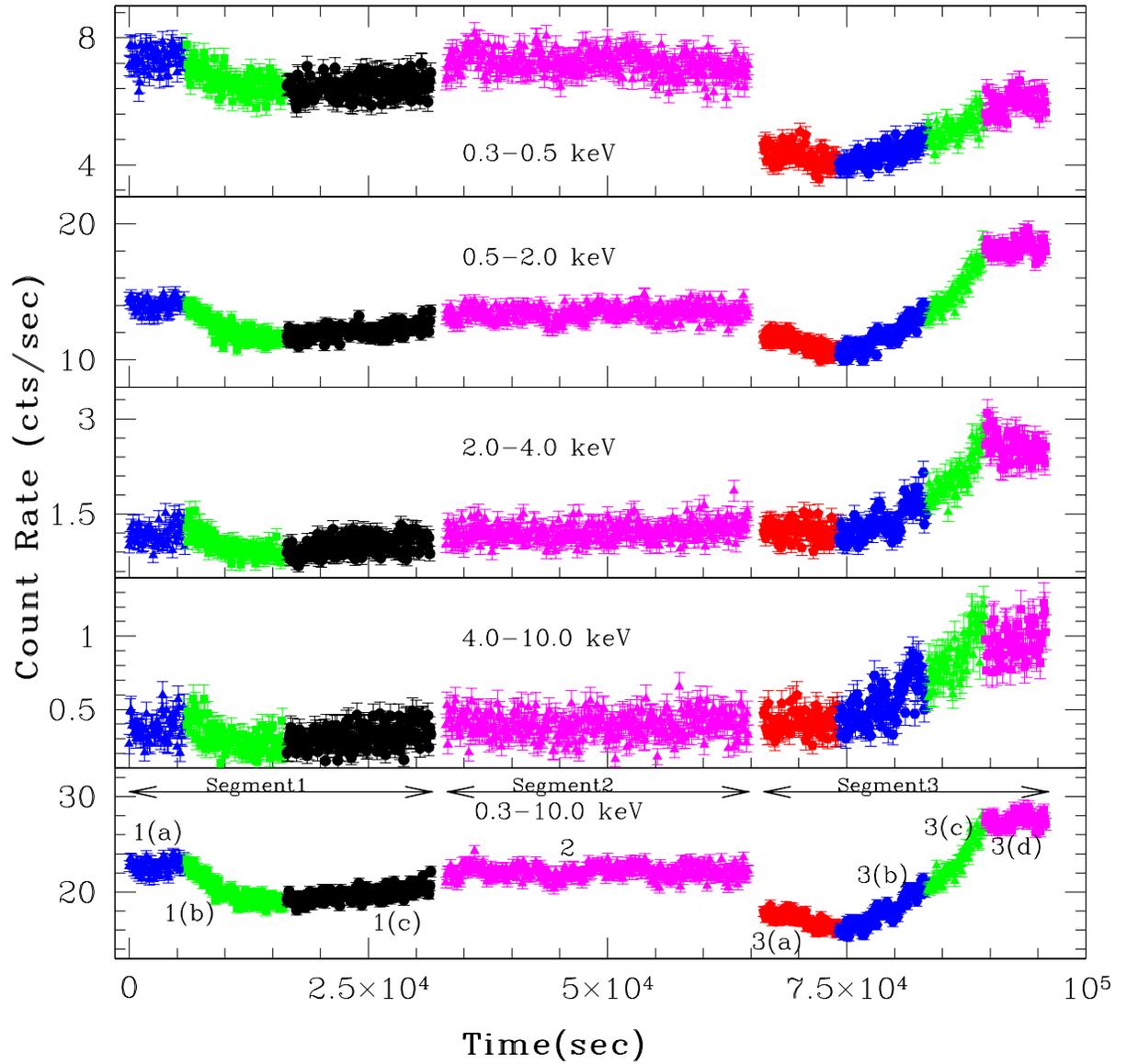}
\caption{The background subtracted X-ray light curve of PKS 2155$-$304 in different energy bands.}
\label{fig:1}
\end{center}
\end{figure}
\newpage
\begin{figure*}
\epsfig{figure= 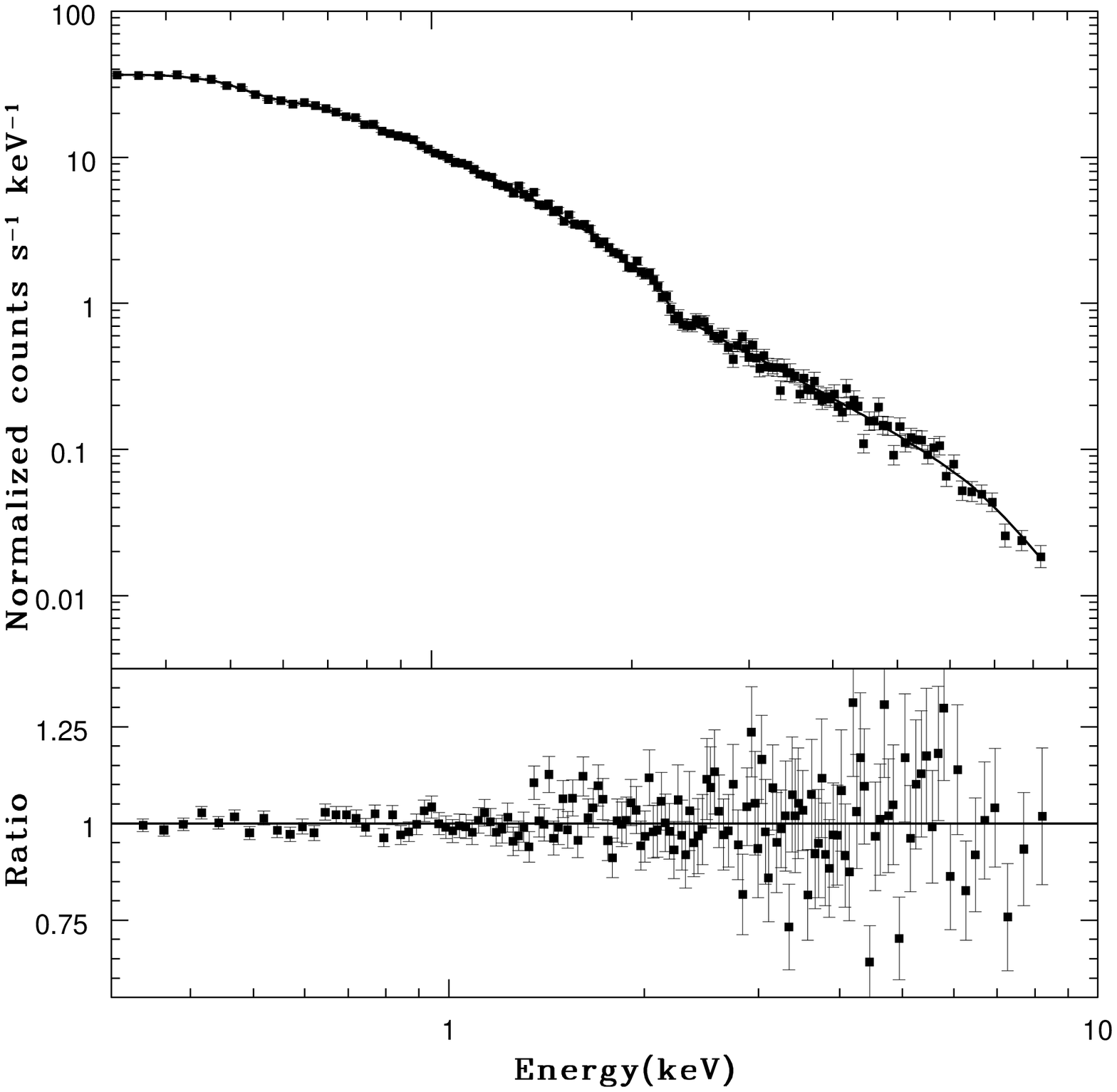,height=2.0in,width=3.1in,angle=0}
\epsfig{figure= 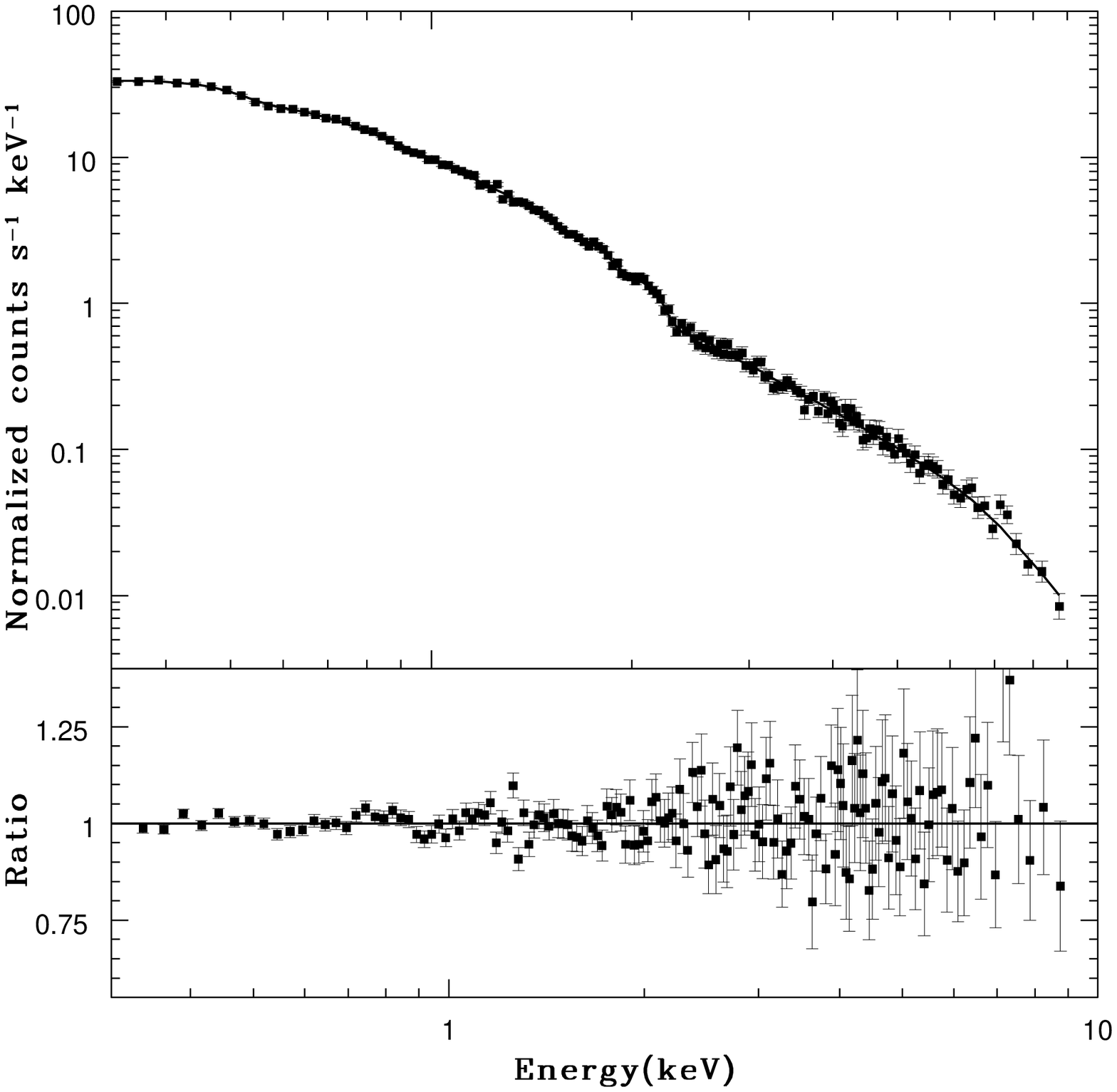,height=2.0in,width=3.1in,angle=0}
\epsfig{figure= 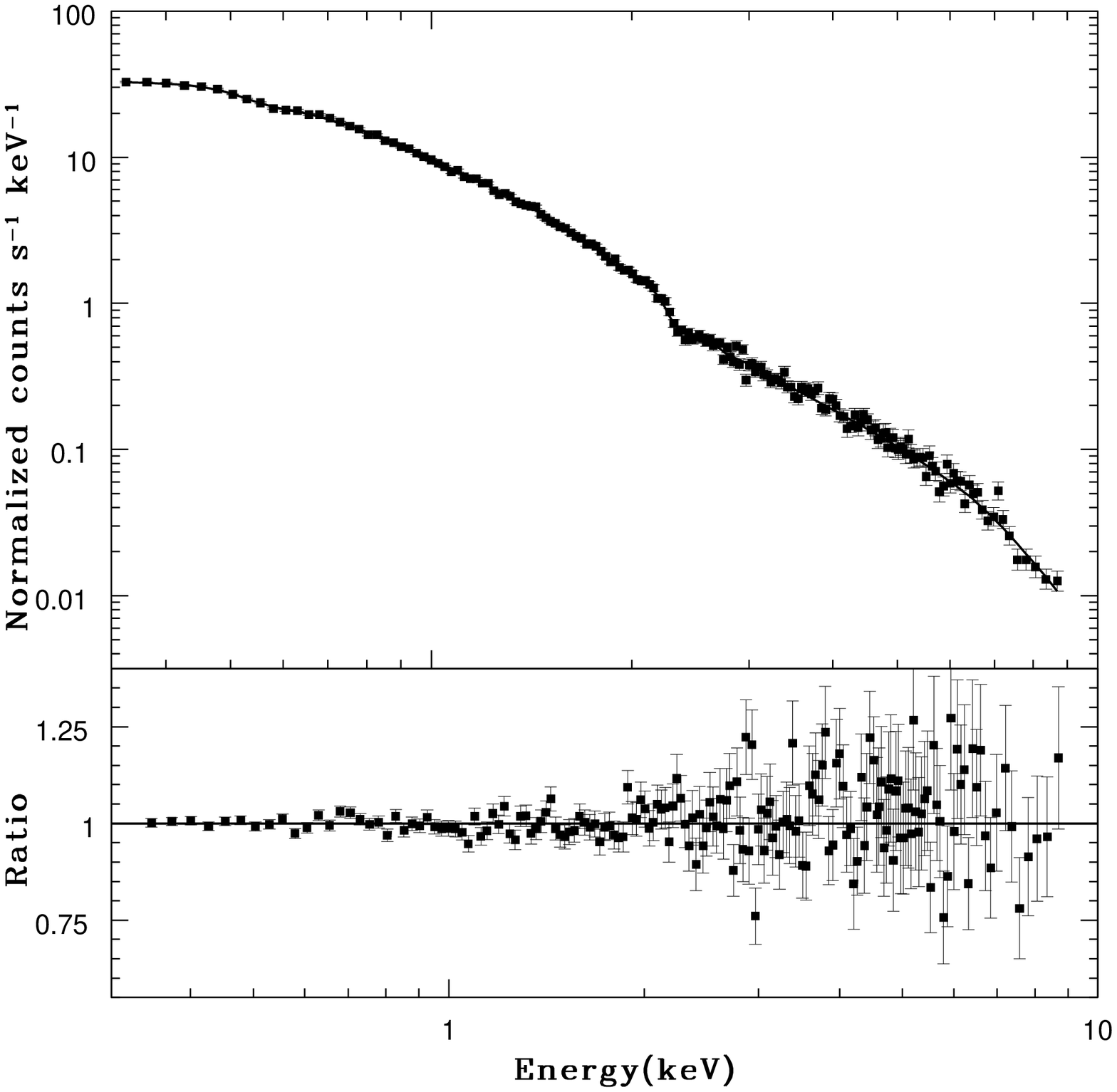,height=2.0in,width=3.1in,angle=0}
\epsfig{figure= 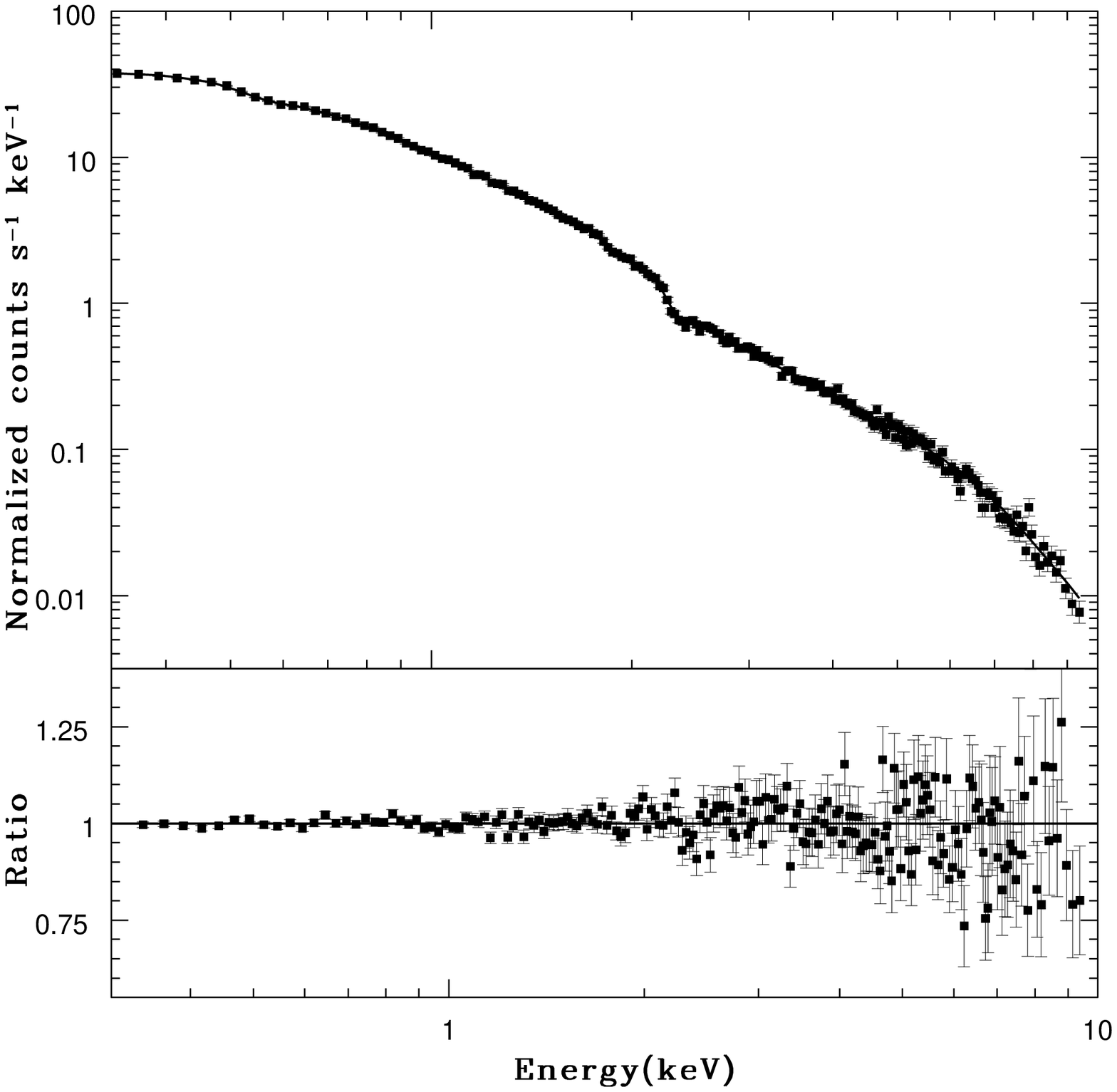,height=2.0in,width=3.1in,angle=0}
\epsfig{figure= 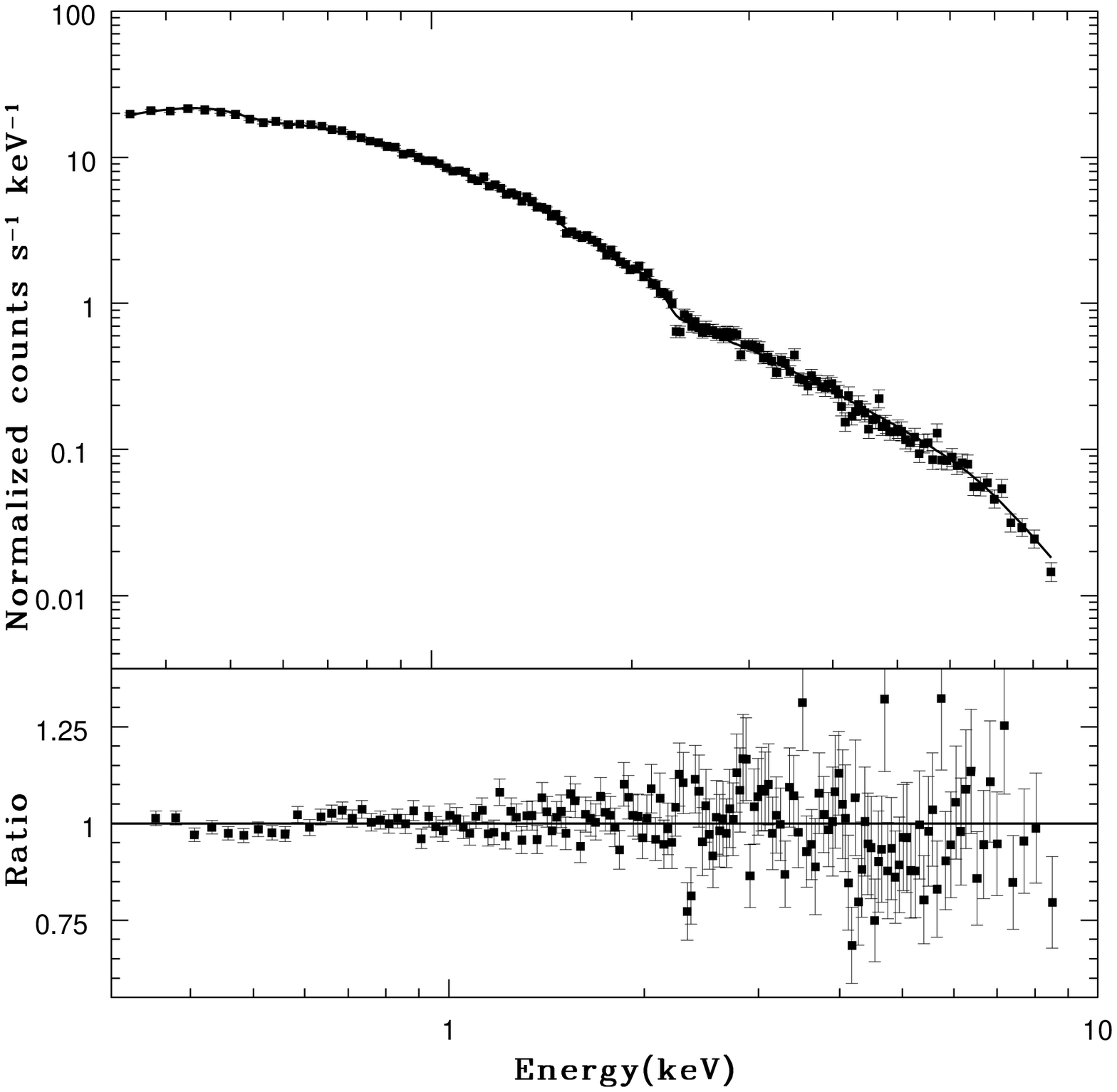,height=2.0in,width=3.1in,angle=0}
\epsfig{figure= 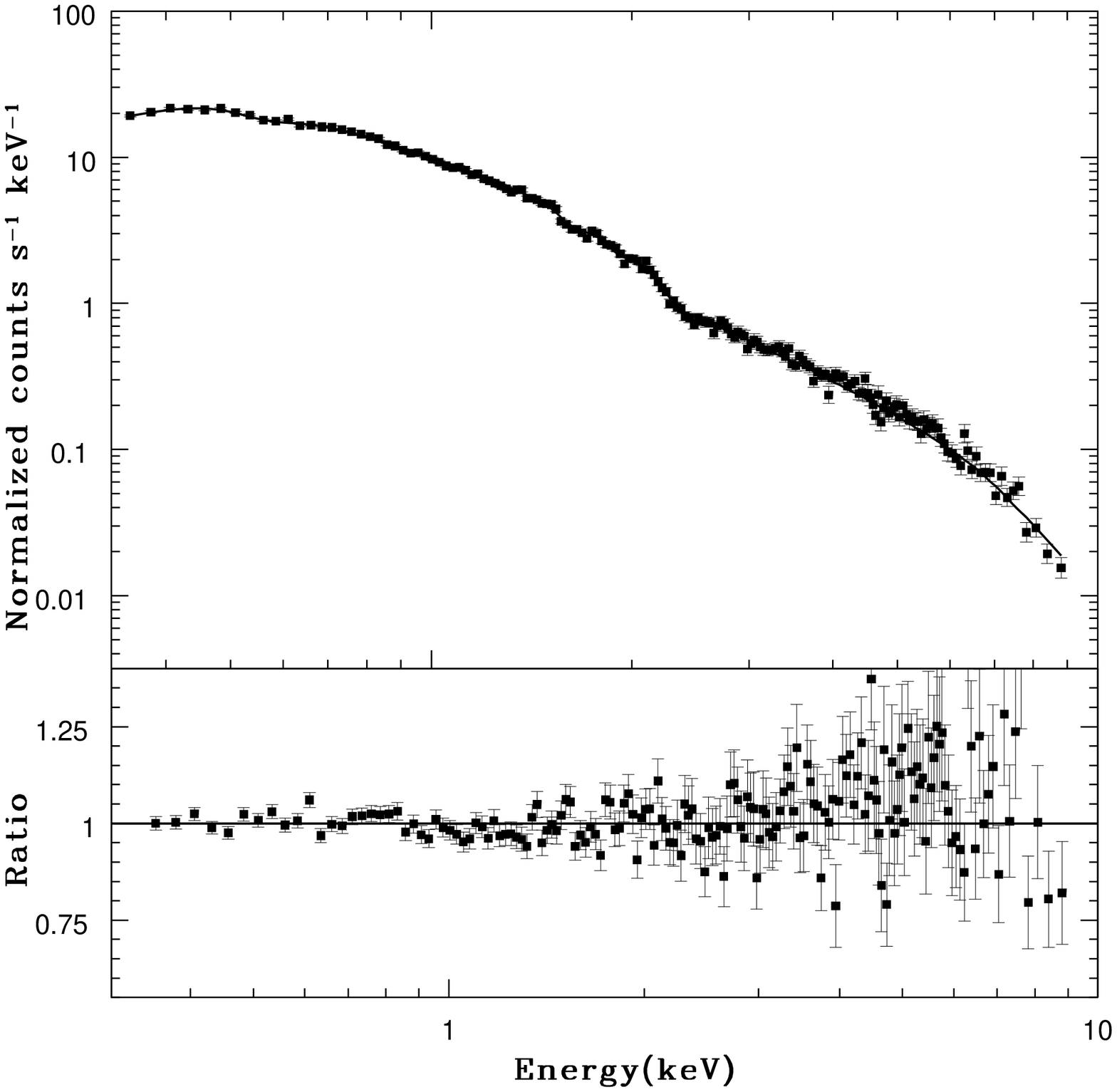,height=2.0in,width=3.1in,angle=0}
\epsfig{figure= 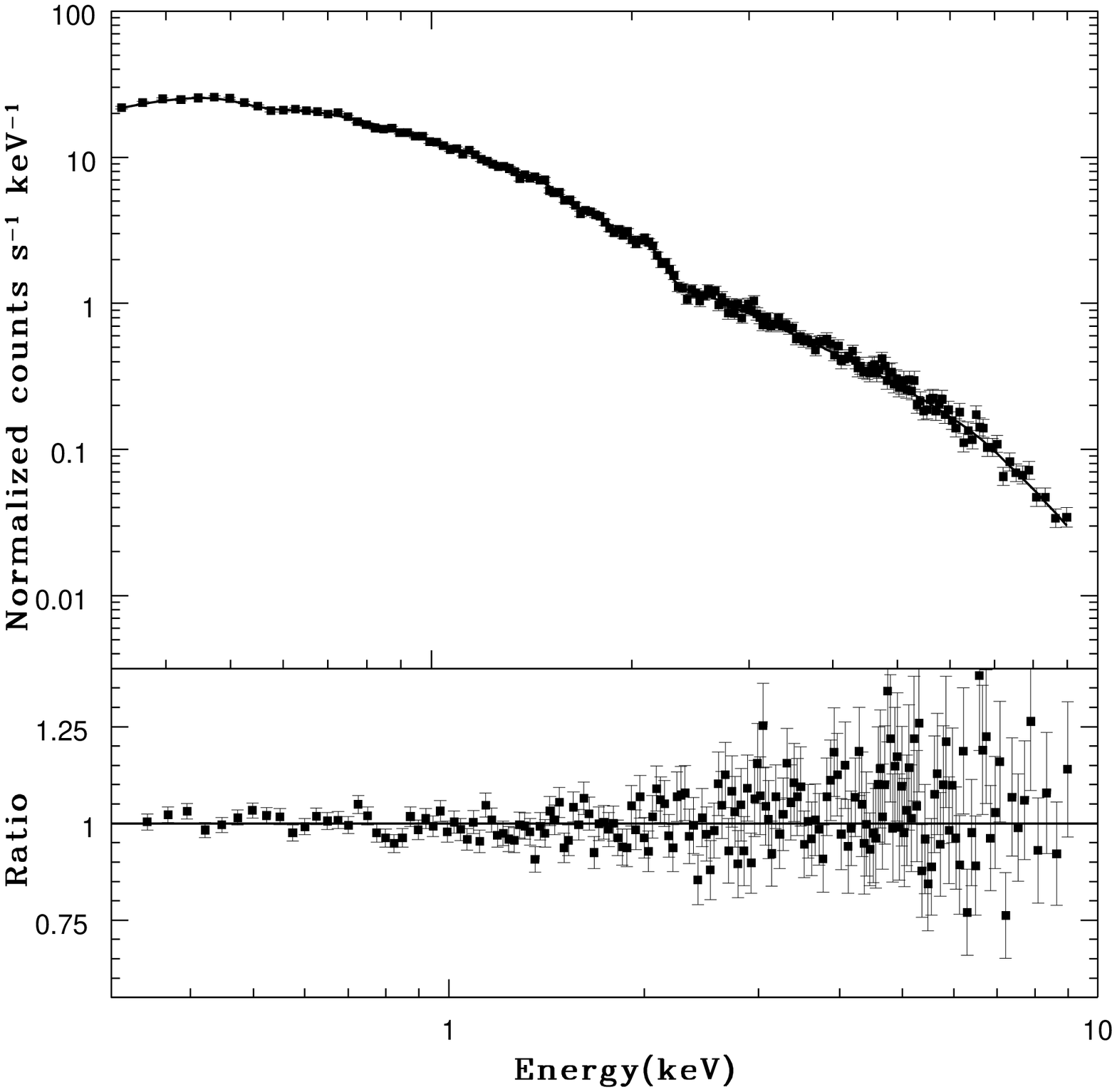,height=2.0in,width=3.1in,angle=0}
\epsfig{figure= 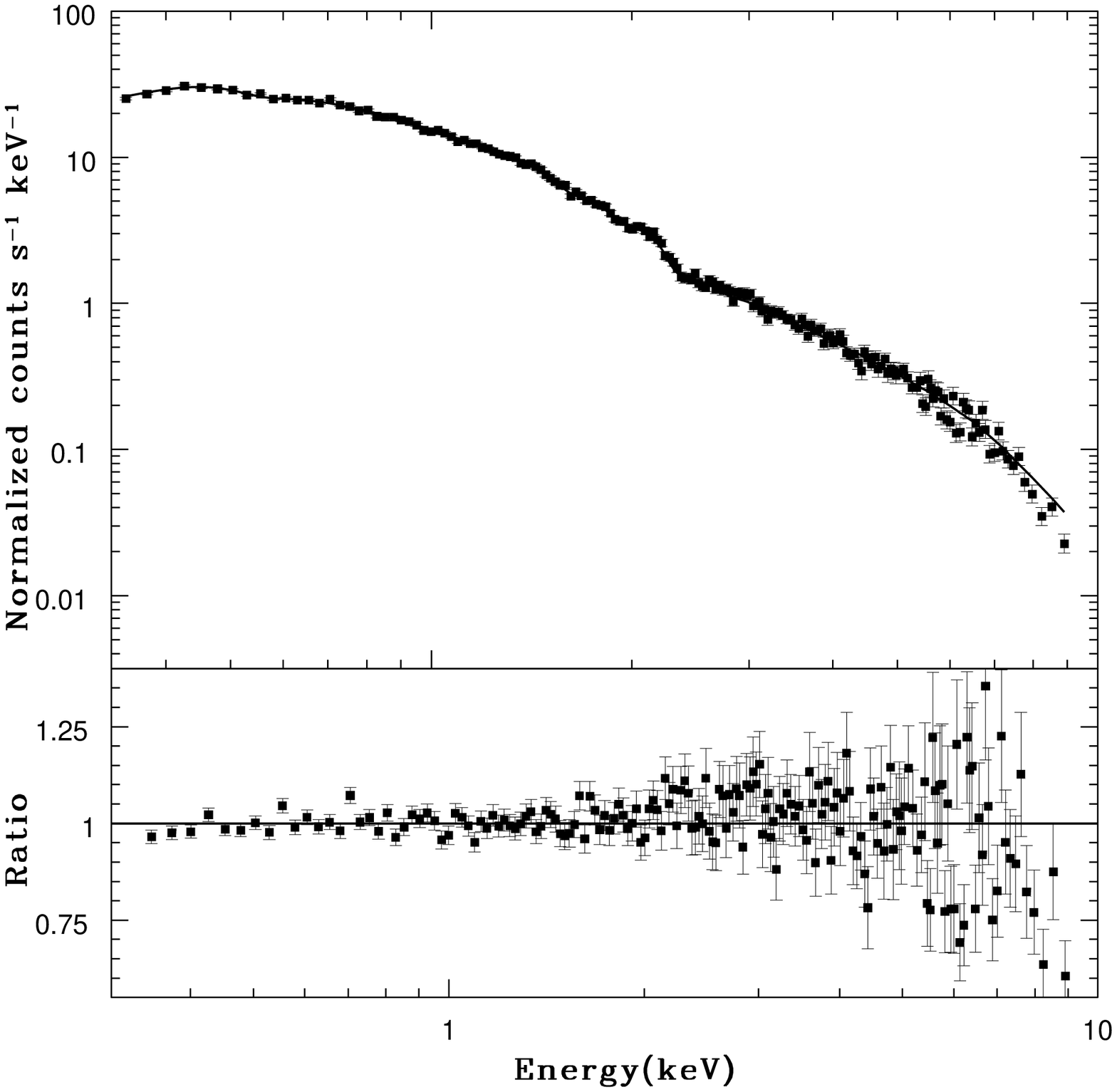,height=2.0in,width=3.1in,angle=0}
\caption{The pn count rate spectra corresponding to the various segments of observation ID 0124930501 
with best fitting simple power-law models and data-to-model ratios. Top row (left to right) have sub-segment 1(a) \& 1(b);
second row have 1(c) \& 2; third row 3(a)\& 3(b) and the last row 3(c)\& 3(d).} 
\label{fig:2}
\end{figure*}
\begin{figure*}
\begin{center}
\includegraphics[width=0.95\textwidth,angle=0]{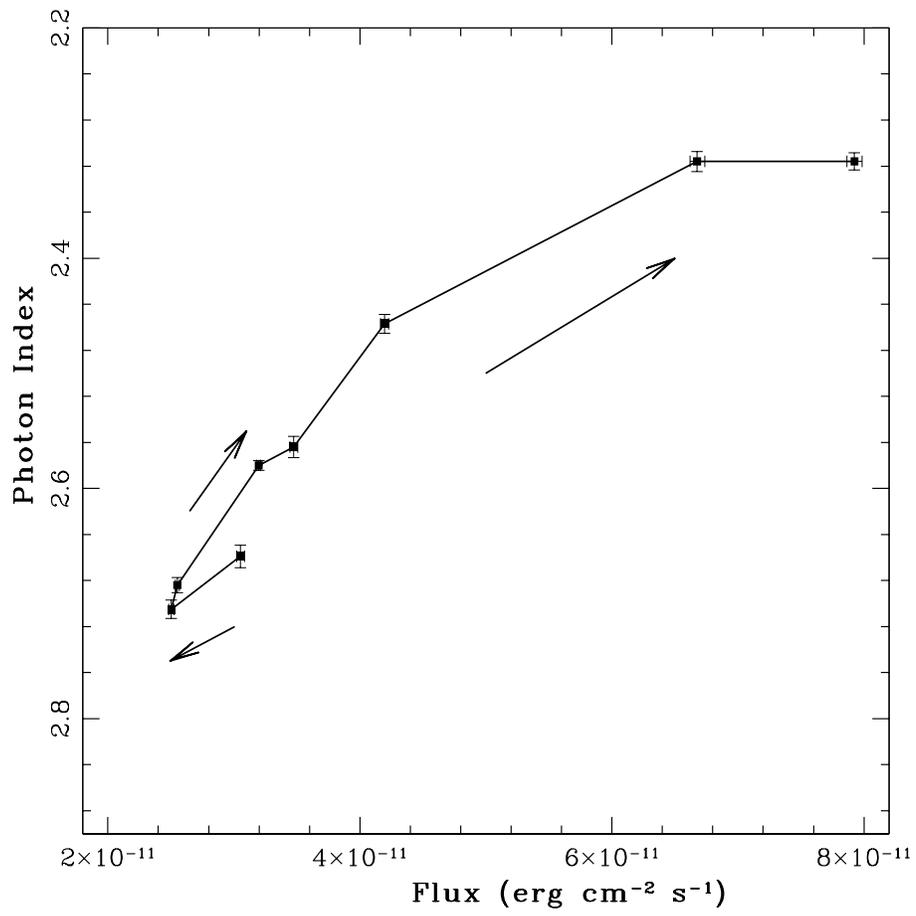}
\caption{Photon-index versus the 2.0-10.0 keV flux}
\label{fig:3}
\end{center}
\end{figure*}

\begin{figure*}
\includegraphics[trim=0.7cm 6.0cm 0.0cm 2.5cm, clip=true, scale=0.95]{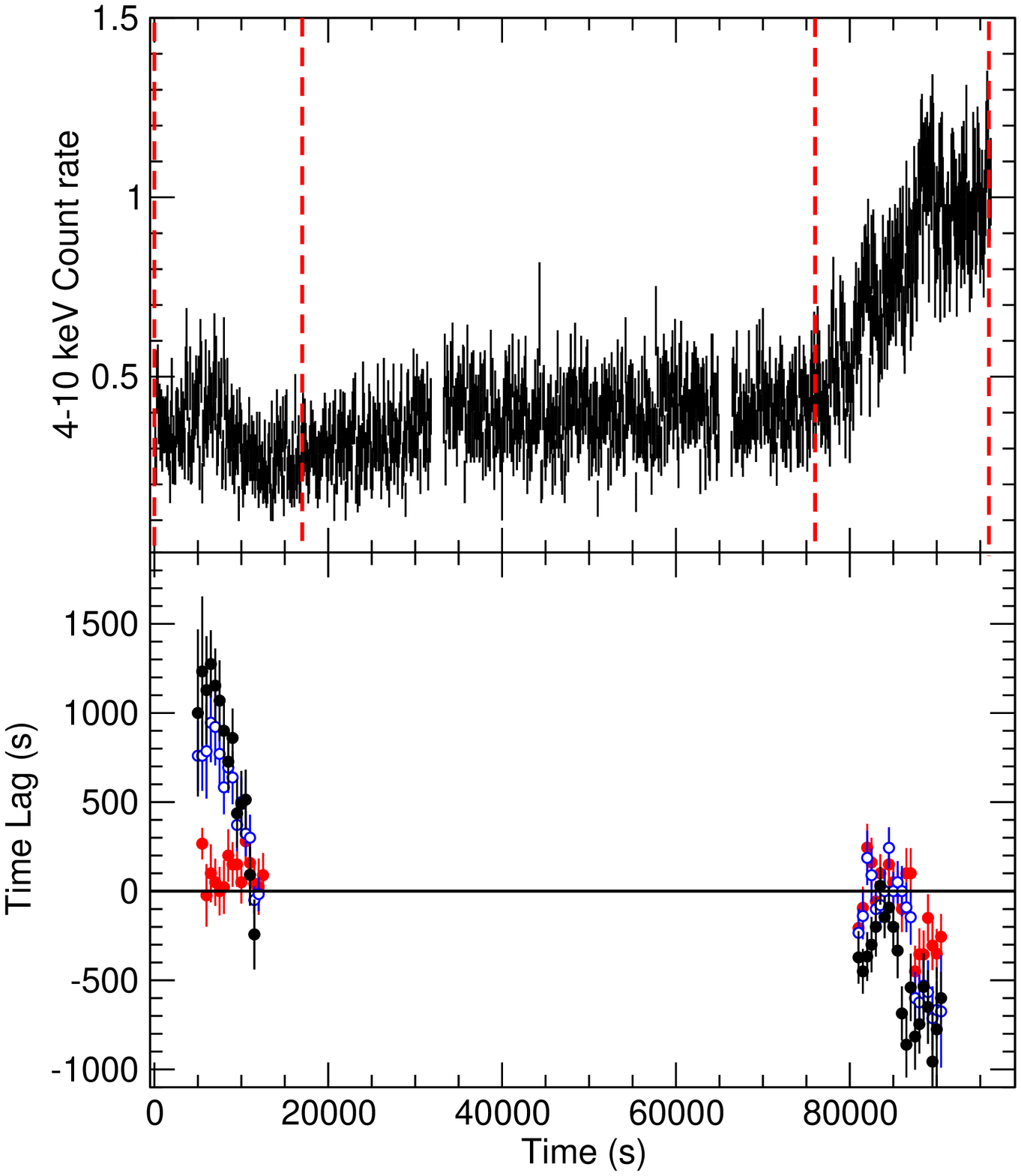}
\caption{The 4--10 keV band light curve (top panel) together with the temporal variation of  the delays between the 0.3--0.5
keV band light curve and the 0.5--2, 2--4 and 4-10 keV band light curves, in the bottom panel (red filled circles, blue open
circles and black field circles, respectively).}
\label{fig:4}
\end{figure*}
\end{document}